# Laser Cooling of Nuclear Magnons

Haowei Xu[1], Guoqing Wang[1,2], Changhao Li[1,2], Hua Wang[1], Hao Tang[3],

Ariel Rebekah Barr[3], Paola Cappellaro[1,2,4, †], and Ju Li[1,3, ‡]

[1] Department of Nuclear Science and Engineering, Massachusetts Institute of Technology, Cambridge, Massachusetts 02139, USA

[2] Research Laboratory of Electronics, Massachusetts Institute of Technology, Cambridge, MA 02139, USA

[3] Department of Materials Science and Engineering, Massachusetts Institute of Technology, Cambridge, Massachusetts 02139, USA

[4] Department of Physics, Massachusetts Institute of Technology, Cambridge, MA 02139, USA

* Corresponding authors: † pcappell@mit.edu,  ‡ liju@mit.edu

**Abstract**

The initialization of nuclear spin to its ground state is challenging due to its small energy scale compared with thermal energy, even at cryogenic temperature. In this Letter, we propose an opto-nuclear quadrupolar effect, whereby two-color optical photons can efficiently interact with nuclear spins. Leveraging such an optical interface, we demonstrate that nuclear magnons, the collective excitations of nuclear spin ensemble, can be cooled down optically. Under feasible experimental conditions, laser cooling can suppress the population and entropy of nuclear magnons by more than two orders of magnitude, which could facilitate the application of nuclear spins in quantum information science.

**Introduction.** Physical qubit platforms are one of the foundations of quantum information science and technology. Nuclear spins have long been perceived as ideal quantum information carriers, thanks to their robustness against environmental perturbations and unparalleled coherence time [1,2]. However, the application of nuclear spins is hindered by several challenges, one of which is the initialization problem – For a typical nuclear spin under a 1 T magnetic field, a 99% initialization fidelity by thermal equilibration requires a demanding temperature below 0.1 mK. The initialization of the nuclear spins can be facilitated by the hyperfine interaction with electron spins, using e.g., dynamic nuclear polarization [3] or optical orientation [4]. But the necessity of ancillary electrons engenders other shortcomings, such as limited applicability only in systems with non-zero electron spins and shortened nuclear spin coherence time [5,6].

Laser cooling of (quasi)-particles, including neutral atoms [7], mechanical modes [8–11], semiconductors [12], and electron magnons [13], has witnessed great success. Optical lasers have also been used to initialize qubit systems, such as electron and nuclear spins (indirectly via the hyperfine interaction) in nitrogen-vacancy centers [14]. If nuclear spins can be cooled down and initialized optically, their applications would be significantly facilitated. However, there is a lack of effective optical interfaces to nuclear spins without electron spins.

In this work, we first introduce the opto-nuclear quadrupolar (ONQ) effect, whereby two-color photons can efficiently interact with nuclear spins without the need for ancillary electron spins. Then we describe the properties of nuclear magnons (NMs), which are the collective excitations of a nuclear spin ensemble (NSE) in crystalline solids such as zinc blende GaAs (zbGaAs) [15–18] and have an exceptionally low decay rate down to ∼ 0.1 kHz. As the ONQ coupling strength between optical photons and NMs scales with the number of nuclear spins as $\sqrt{N}$, the ONQ effect is suitable for controlling large NSE. Taking advantage of these properties, we demonstrate the laser cooling of the NM via the ONQ effect. From an initial temperature of mK obtainable in dilute refrigerators [19], the population and the entropy of the NM can be simultaneously reduced by more than two orders of magnitude under feasible experimental conditions.

**Opto-nuclear quadrupolar effect.** The Hamiltonian of a nucleus with spin $I > \frac{1}{2}$ is $H_n = \gamma_n \mathcal{B} \cdot I + I \cdot \mathcal{Q} \cdot I = \gamma_n \sum_i \mathcal{B}_i I_i + \sum_{ij} \mathcal{Q}_{ij} I_i I_j$, where the first and second terms are the nuclear magnetic (Zeeman) and nuclear electric quadrupole interactions, respectively. $\gamma_n$ is the nuclear gyromagnetic ratio, $\mathcal{B}$ is the magnetic field, $I$ is the nuclear spin operator, and $i, j = x, y, z$ are Cartesian indices. The Zeeman interaction comes from the nuclear magnetic dipole. Then in non-spherical nuclei, an electric quadrupole moment $q$ also arises as the leading order electric moment when one performs the multi-pole expansion (the nuclear electric dipole is zero because of inversion symmetry, see e.g., Chapter 3 in Ref. [20]). The interaction between the nuclear electric quadrupole moment and the electric field gradient (EFG) at the site of the nucleus leads to the nuclear quadrupole interaction $\mathcal{Q}_{ij} \equiv \frac{eq\mathcal{V}_{ij}}{2I(2I-1)}$, where $\mathcal{V}_{ij}$ is the EFG operator.

Traditional techniques for controlling nuclear spins (e.g., nuclear magnetic resonance) rely on modulating the Zeeman interaction using microwave magnetic fields. It is also possible to control nuclear spins by modulating the EFG through electric interaction with the nuclear spin. Particularly, one can use external electric field(s) to drive the orbital motion of electrons, so that there is a change $\Delta \mathcal{V}$ in the EFG generated by electrons. Under two-color electric fields $\mathcal{E}_{p(q)}(t) = \mathcal{E}_{p(q)} e^{i\omega_{p(q)} t}$, the electron cloud oscillates in real space with a frequency $\omega_p - \omega_q$ (Figure 1a). Consequently, the EFG generated by electrons and thus the nuclear electric quadrupole interaction will also have an oscillating part with frequency $\omega_p - \omega_q$, which can match nuclear spin energies. This is what we call the ONQ effect. The ONQ effect is a cousin process of Raman scattering or difference frequency generation (DFG) [21]. In Raman (DFG), the oscillation of electrons leads to the emission of phonons (photons) at the difference-frequency $\omega_p - \omega_q$; In ONQ, the oscillation of electrons results in the oscillations of the nuclear electric quadrupole interaction at the difference-frequency.

Formally, the oscillating nuclear quadrupole interaction can be expressed as

$$H_{\mathrm{ONQ}} = \mathcal{D}_{ij}^{pq}(\omega_p - \omega_q; \omega_p, -\omega_q) \mathcal{E}_p(\omega_p) \mathcal{E}_q(-\omega_q) I_i I_j e^{i(\omega_p - \omega_q)t} + h.c. , \qquad (1)$$

where $h.c.$ stands for Hermitian conjugate. Terms with frequencies $\omega_p$, $\omega_q$, and $\omega_p + \omega_q$ are far off-resonance with nuclear spin dynamics and are omitted. $\mathcal{D}_{ij}^{pq} \equiv \frac{\partial^2 Q_{ij}}{\partial \mathcal{E}_p \partial \mathcal{E}_q}$ is the second-order response function of the quadrupole tensor. In the single-particle approximation, one has [22]

$$\begin{aligned}&\mathcal{D}_{ij}^{pq}(\omega_p - \omega_q; \omega_p, -\omega_q) \\ &= \frac{e^3 q}{2I(2I-1)} \sum_{mnl} \frac{[\mathcal{V}_{ij}]_{mn}}{E_{mn} - \hbar(\omega_p - \omega_q)} \times \left\{ \frac{f_{lm}[r_p]_{nl}[r_q]_{lm}}{E_{ml} - \hbar\omega_p} - \frac{f_{nl}[r_q]_{nl}[r_p]_{lm}}{E_{ln} - \hbar\omega_p} \right\} + (p \leftrightarrow q), \end{aligned} \qquad (2)$$

where $(p \leftrightarrow q)$ indicates the exchange of the $p$ and $q$ subscripts, which symmetrizes the $\omega_p$ and $\omega_q$-fields. $m, n, l$ label the electronic states, $E_{mn}$ and $f_{mn}$ are the energy and occupation differences between two electronic states $|m\rangle$ and $|n\rangle$. Meanwhile, $[r_i]_{mn} \equiv \langle m | r_i | n \rangle$ is the position operator, and $[\mathcal{V}_{ij}]_{mn} = \frac{e}{4\pi\varepsilon_0} \langle m | \frac{3 r_i r_j - \delta_{ij} r^2}{r^5} | n \rangle$ is the EFG operator of the electrons, with $\varepsilon_0$ being the vacuum permittivity.

Notably, electron spin operators do not explicitly appear in Eq. (2), corroborating that the ONQ effect does not need ancillary electron spin. Besides, the light frequency $\omega_{p(q)}$ only appears in the denominators. Hence, the $\mathcal{D}$ tensor is insensitive to $\omega_{p(q)}$ when they are not close to the electron bandgap $E_g$, leading to flexibility in choosing $\omega_{p(q)}$; Moreover, all electrons contribute to the ONQ response, as indicated by the summation over $(m, n, l)$ indices. When $\omega_{p(q)} > E_g$, electrons can do resonant interband transitions. When $\omega_{p(q)} < E_g$, the electron interband transitions are virtual. We will consider $\omega_{p(q)} < E_g$, to avoid resonant one-photon absorptions (Section 3 of Ref. [23], which also contains Refs. [1,2,24–64]).

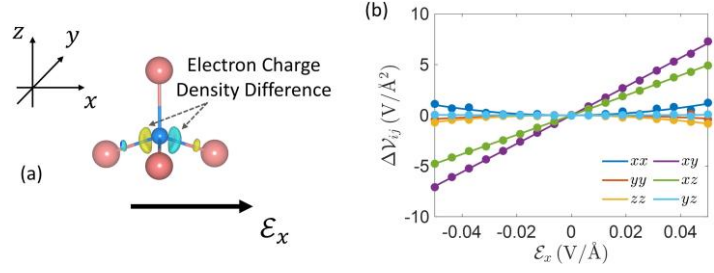

**Figure 1** The ONQ effect in zinc-blende GaAs. **(a)** Yellow (green) bubbles denote positive (negative) changes in electron charge density when an electric field $\mathcal{E}_x$ is applied. Pink (blue) spheres are Ga (As) atoms. **(b)** $\Delta \mathcal{V}_{ij}$ at the site of As nuclei as a function of $\mathcal{E}_x$.

**Magnitude of the $\mathcal{D}$ tensor.** For an order-of-magnitude estimation of the $\mathcal{D}$ tensor, we use $\left\langle m \left| \frac{3r_i r_j - \delta_{ij} r^2}{r^5} \right| n \right\rangle \approx \frac{1}{a_0^3}$ and $[r_i]_{mn} \approx a_0$ in Eq. (2). Here $a_0$ is the Bohr radius, which is also approximately half the bond length in typical materials. In addition, we only consider the $(m, n, l)$ pair that satisfies $E_{mn} = E_{ml} = E_g$, which makes the major contribution to $\mathcal{D}$ when $\omega_{p(q)} < E_g$. Then, one has $\mathcal{D} \sim \frac{g_S}{2I(2I-1)} \frac{e^4 q}{4\pi\varepsilon_0 a_0} \frac{1}{E_g(E_g - \omega_p)}$ with $g_S = 2$ the electron spin degeneracy. As an example, this estimate yields $\mathcal{D} \sim 0.24 \times \frac{2\pi \cdot \text{Hz}}{(\text{MV/m})^2}$ for $^{75}$As nuclei in zbGaAs when $E_g - \omega_p = 0.2$ eV. The $\mathcal{D}$ tensor can also be evaluated using density functional theory (DFT, Section 4.1 of Ref. [23]). We apply a static electric field $\mathcal{E}$ and calculate the change in EFG $\Delta \mathcal{V}$. Then $\mathcal{D}$ in the static limit ($\omega_p = \omega_q = 0$) can be obtained by fitting the $\Delta \mathcal{V}$ - $\mathcal{E}$ curve (Figure 1b, 1 V/Å = $10^4$ MV/m), yielding $\mathcal{D}(0; 0,0) \approx 0.20 \times \frac{2\pi \cdot \text{Hz}}{(\text{MV/m})^2}$ for $^{75}$As nuclei in zbGaAs, in reasonable agreement with the analytical estimate above. Notably, due to the tetrahedral symmetry of zbGaAs, one has $\mathcal{Q} = 0$ when $\mathcal{E} = 0$. However, $\mathcal{D}$ is non-zero. The validity of the estimation of the $\mathcal{D}$ tensor is assessed in Section 4.2 of Ref. [23]. We will adopt $\mathcal{D} = 0.2 \times \frac{2\pi \cdot \text{Hz}}{(\text{MV/m})^2}$ hereafter. For a single nuclear spin (Section 2.5 of Ref. [23]), the ONQ coupling strength is only 20 Hz when $\mathcal{E}_p = \mathcal{E}_q = 10$ MV/m. Fortunately, as we will show later, the collective ONQ coupling of an NSE can be boosted by a $\sqrt{N}$ factor. Hence, we will focus on NSE hereafter.

**Properties of nuclear magnons.** In analogy with electronic spin magnons [64,65], nuclear spin magnons are collective excitation modes of nuclear spins. For brevity, we assume the nuclei are of the same species. The Hamiltonian of an NSE is $\mathcal{H} = \sum_\alpha (\gamma_n \boldsymbol{I}^\alpha \cdot \boldsymbol{B} + \boldsymbol{I}^\alpha \cdot \boldsymbol{Q} \cdot \boldsymbol{I}^\alpha) + \sum_{\alpha\beta} \boldsymbol{I}^\alpha \cdot \boldsymbol{\mathcal{J}}^{\alpha\beta} \cdot \boldsymbol{I}^\beta$, where $\mathcal{J}^{\alpha\beta}$ describes the interaction between two nuclear spins $\alpha$ and $\beta$. The spin operators $\boldsymbol{I}^\alpha$ can be converted to NM creation (annihilation) operators $a_{\boldsymbol{k}}^\dagger$ ($a_{\boldsymbol{k}}$) with $\boldsymbol{k}$ the wavevector (Section 1 of Ref. [23]). Figure 2a shows a semi-classical one-dimensional illustration of the NM. Each nuclear spin precesses around its ground state, and the phase of the precession is $e^{i\boldsymbol{k}\cdot\boldsymbol{r}_\alpha}$ with $\boldsymbol{r}_\alpha$ the location of the $\alpha$-th nucleus, so the wavelength is $\lambda = \frac{2\pi}{|\boldsymbol{k}|}$. This resembles the phonons, whereby the atomic vibrations have a $e^{i\boldsymbol{k}\cdot\boldsymbol{r}_\alpha}$ phase factor.

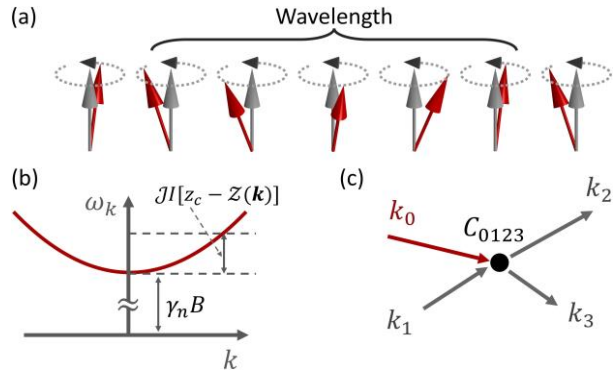

**Figure 2 (a)** A semi-classical one-dimensional illustration of the NM mode. **(b)** Band dispersion of the NMs (not to scale). One has $\gamma_n \mathcal{B} \gg \mathcal{J}I[z_c - \mathcal{Z}(\boldsymbol{k})]$. **(c)** Illustration of the four-NM scattering process.

In the basis of $a_{\boldsymbol{k}}^\dagger$ ($a_{\boldsymbol{k}}$), $\mathcal{H}$ can be decomposed as $\mathcal{H} = \mathcal{H}^{(2)} + \mathcal{H}^{(3)} + \mathcal{H}^{(4)} + \cdots$, where $\mathcal{H}^{(\zeta)}$ contains $\zeta$ NM annihilation/creation operators (Section 1.2 of Ref. [23]). The quadratic term, $\mathcal{H}^{(2)} = \sum_{\boldsymbol{k}} \omega_{\boldsymbol{k}} a_{\boldsymbol{k}}^\dagger a_{\boldsymbol{k}}$, sets the NM frequency $\omega_{\boldsymbol{k}}$. Higher-order terms such as $\mathcal{H}^{(3)}$ and $\mathcal{H}^{(4)}$, which arise from the $\boldsymbol{\mathcal{J}}$ and $\boldsymbol{Q}$ terms, correspond to multi-NM interactions and lead to the relaxation of NMs [24]. We will set $\boldsymbol{Q} = 0$, suitable for GaAs. We also assume a nearest-neighbor Heisenberg interaction $\mathcal{J}_{ij}^{\alpha\beta} = \mathcal{J}\delta_{\langle\alpha\beta\rangle}\delta_{ij}$, where $\delta_{ij}$ is the Kronecker delta, and $\delta_{\langle\alpha\beta\rangle}$ enforces $\alpha$ and $\beta$ to be nearest neighbors. These approximations would not change the order-of-magnitude of the results below (Section 1.1 of Ref. [23]). Then, one has

$$\omega_{\boldsymbol{k}} = \gamma_n \mathcal{B} + \mathcal{J}I[z_c - \mathcal{Z}(\boldsymbol{k})], \tag{3}$$

where $z_c$ is the coordination number, while $\mathcal{Z}(\boldsymbol{k})$ depends on the lattice structure and is on the order of unity (Section 1.1 of Ref. [23]). Notably, the NM bandwidth ($\mathcal{J}I \sim$ kHz) is much smaller than $\gamma_n \mathcal{B}$ (above MHz when $\mathcal{B}$ is on the order of Tesla), and thus one has $\omega_{\boldsymbol{k}} \approx \omega_0 \equiv$ $\gamma_n \mathcal{B}$. The subscript 0 denotes the near-$\Gamma$-point NM mode ($\boldsymbol{k}_0 \approx 0$), which can interact with optical photons and will be the focus henceforth.

The relaxation rate $\kappa_0$ of the near-$\Gamma$-point NM is a crucial parameter in the laser cooling processes, as we will show below. Due to the small NM bandwidth, three-NM scatterings always violate the conservation of energy, and thus cannot lead to NM relaxation. The leading-order contribution to NM relaxation comes from four-NM scatterings described by $\mathcal{H}^{(4)} = \sum_{0123} C_{0123} a_0 a_1 a_2^\dagger a_3^\dagger + h.c.$ (the $\boldsymbol{k}_0 + \boldsymbol{k}_1 \to \boldsymbol{k}_2 + \boldsymbol{k}_3$ scattering, Figure 2c). Here $l = 1,2,3$ label three other NMs interacting with the near-$\Gamma$-point NM ($l = 0$). The four-NM coupling strength $C_{0123}$ depends on $\mathcal{J}$. Note that $\mathcal{H}^{(4)}$ also contains other terms such as $a_0 a_1^\dagger a_2^\dagger a_3^\dagger$, which are excluded because they violate energy conservation. The relaxation rate due to four-NM scatterings is $\kappa_0^{(4)} \approx \frac{\pi}{2} \left(\frac{3}{4\pi}\right)^{\frac{4}{3}} \frac{\mathcal{J}}{I\hbar} n_0(n_0 + 1)$. Notably, the relaxation rate depends on $\mathcal{J}$ and $I$, which are respectively the inter-nuclear interaction strength and the nuclear angular momentum. Specifically, one has $\kappa_0^{(4)} \lesssim [0.1 \sim 1]$ kHz when $n_0 \sim 1$. .We set the total NM relaxation rate as $\kappa_0 = \kappa_0^{(4)}$ henceforth, as contributions from higher-order terms $\mathcal{H}^{(\zeta > 4)}$ are minor (Section 1.2 of Ref. [23], see also Refs. [60,66,67]).

**ONQ interaction of nuclear magnons.** Next, we discuss the collective ONQ interaction between optical photons and NMs. To achieve a laser cooling effect, the system is put in an optical cavity resonant with the $\omega_q$-photon and is pumped with the $\omega_p$-laser. Hence, we second-quantize the $\omega_q$-photon and treat the $\omega_p$-laser as a classical field. The conservation of energy enforces $\omega_q = \omega_p \pm \omega_0$. Specifically, an optical photon with shifted frequency $\omega_h = \omega_p + \omega_0$ ($\omega_l = \omega_p - \omega_0$) is emitted when an NM is annihilated (created), which can be described by (Section 1.3 of Ref. [23])

$$\mathcal{H}_{\mathrm{ONQ}} = \mathcal{G}_h b_h^\dagger a_0 + \mathcal{G}_l b_l^\dagger a_0^\dagger + h.c., \qquad (4)$$

where $b_{h(l)}^\dagger$ is the creation operator of the $\omega_{h(l)}$-photon, and

$$\mathcal{G}_{h(l)} \equiv g\sqrt{N}\mathcal{E}_p \mathcal{E}_{h(l)}^{\mathrm{zpf}} \qquad (5)$$

is the collective ONQ coupling strength for NMs with $g \sim \mathcal{D}_{ij}^{pq} \approx 0.2 \times \frac{2\pi \cdot \mathrm{Hz}}{(\mathrm{MV/m})^2}$. $\mathcal{E}_{h(l)}^{\mathrm{zpf}}$ is the zero-point field strength of the $\omega_{h(l)}$-photon. Remarkably, $\mathcal{G}_{h(l)}$ is enhanced by a $\sqrt{N}$ factor, similar to the collective coupling between photons and Dicke atomic states or phonons [68,69].

This $\sqrt{N}$ factor indicates that the ONQ effect is suitable for controlling large NSE, which can have sizable interaction with a single cavity photon even if the pumping field $\mathcal{E}_p$ is mild.

**Laser cooling mechanism.** The possible transitions of the NM mode under the $\omega_p$-laser are illustrated in the inset of Figure 3c. Green (red) arrows correspond to the first (second) term in Eq. (4). Efficient laser cooling requires $\mathcal{G}_h \gg \mathcal{G}_l$ [68], which can be realized by using an optical cavity resonant with the $\omega_h$-photon, whereby one has $\mathcal{E}_h^{\text{zpf}} = \sqrt{\frac{\hbar\omega_h}{2\varepsilon_0 V_h}}$ with $V_h$ the mode volume of the $\omega_h$-cavity. The solid green arrows indicate the $\omega_p + \omega_0 \to \omega_h$ process, which annihilates and cools down the NMs. The reverse $\omega_h \to \omega_p + \omega_0$ transition (dashed green arrows) creates NMs and is the back-heating effect. Fortunately, the back-heating can be suppressed by keeping the population of the $\omega_h$-photon small (ideally zero) via a thermal energy much lower than $\omega_h$. This cooling mechanism is similar to the anti-Stokes cooling of phonons [8–11,68].

In the rotating frame of $\omega_p$, the Hamiltonian of the combined system of $\omega_h$-photons and NMs is

$$\mathcal{H}_C = \omega_0 a_0^\dagger a_0 + (\omega_h - \omega_p) b_h^\dagger b_h + \left(\mathcal{G}_h b_h^\dagger a_0 + h.c.\right). \tag{6}$$

We further assume $\omega_h = 1\,\text{eV}$ and $\frac{N}{V_h} \approx \rho_n$. Here $\rho_n$ is the number density of the nuclear spins, which basically is also the number density of atoms. In solid-state systems, $\rho_n$ can reach $10^{28} \sim 10^{29}\,\text{m}^{-3}$ (for example, one has $\rho \approx 4.2 \times 10^{28}\,\text{m}^{-3}$ for As in zbGaAs). For clarity, we will use $\rho_n = 10^{28}\,\text{m}^{-3}$ hereafter. This leads to $\mathcal{G}_h[\text{kHz}] \approx 1.9 \times \mathcal{E}_p[\text{MV/m}]$, that is, a $1\,\text{MV/m}$ pumping field (intensity $\approx 1.3\,\text{mW} \cdot \text{μm}^{-2}$) leads to a collective ONQ coupling strength of $1.9\,\text{kHz}$. In practice, much higher laser power above tens of Watt can be obtained [70]. Note that $\mathcal{G}_h$ scales as $\sqrt{N/V_h}$, which is the achievable number density $\rho_n$ in the cavity volume $V_h$ (Section 3.1 of Ref. [23]).

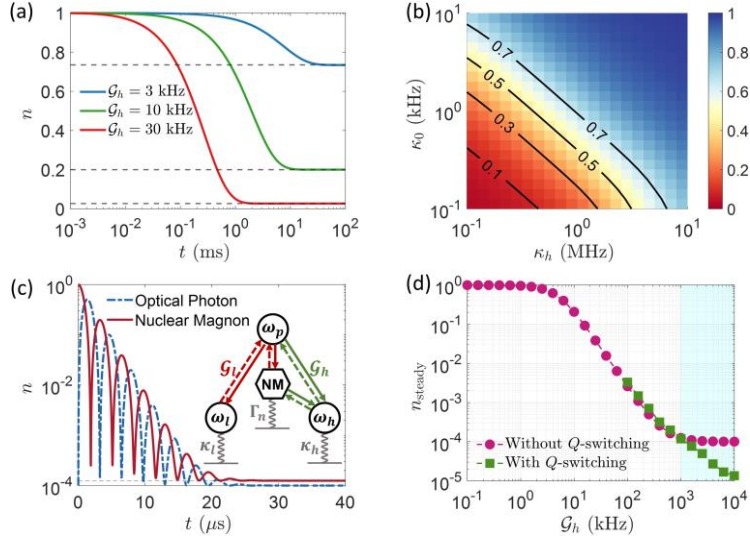

**Figure 3** Laser cooling dynamics. The initial NM population is $n_{\text{th}} = 1$. **(a)** Time evolution of $n_0$ in the weak-coupling regime. **(b)** $n_0^{\text{steady}}$ as a function of $\kappa_0$ and $\kappa_h$ in the weak-coupling regime. $\mathcal{G}_h = 10$ kHz. **(c)** Time evolution of $n_0$ and $n_h$ in the strong coupling regime. $\mathcal{G}_h = 1$ MHz. Inset of (c) shows possible transitions of the NMs. Circles denote optical photons/lasers with frequencies marked inside. The hexagon denotes the NM. Green and red arrows denote ONQ transitions. Grey wavy lines denote coupling with the heat bath. **(d)** $n_0^{\text{steady}}$ as a function of $\mathcal{G}_h$. The red (green) curve denotes laser cooling without (with) $Q$-switching of the optical cavity. The cyan-shaded area corresponds to the strong-coupling regime. In (a, c, d), $\kappa_0 = 0.1$ kHz and $\kappa_h = 1$ MHz are used.

**Laser cooling dynamics**. To demonstrate the laser cooling dynamics, we numerically solve the master equation

$$\frac{\partial \rho}{\partial t} = i[\rho, \mathcal{H}_C] + \kappa_h \xi[b_h]\rho + \kappa_0(n_{\text{th}} + 1)\xi[a_0]\rho + \kappa_0 n_{\text{th}} \xi[a_0^\dagger]\rho, \qquad (7)$$

where $\rho$ is the density matrix of the total system. The Lindblad operator for a given operator $o$ is $\xi(o) = o\rho o^+ - \frac{1}{2}(o^+ o\rho + \rho o^+ o)$. The dissipation rate of the $\omega_h$-photon is $\kappa_h = \frac{\omega_h}{Q_h}$ with $Q_h$ the quality factor of the $\omega_h$-cavity. $n_{\text{th}} = \left[\exp\left(\frac{\omega_0}{k_B T}\right) - 1\right]^{-1}$ is the thermal population of the NM mode at temperature $T$. Considering that $\omega_0$ can be tens of MHz under a magnetic field of 1Tesla, while $T$ can reach mK in a dilution refrigerator, we fix $n_{\text{th}} = 1$ hereafter. It is also possible to start from a higher temperature and larger $n_{\text{th}}$, but this would make $\kappa_0$ larger and the laser cooling less efficient. The thermal population of the $\omega_h$-photon is ignored since $\omega_h \gg k_B T$.

The laser cooling behavior is characterized by two parameters $\frac{\mathcal{G}_h}{\kappa_0}$ and $\frac{\mathcal{G}_h}{\kappa_h}$. $\kappa_0$ is usually in the sub-kHz range, while $\mathcal{G}_h$ can be well above 1 kHz. Hence, we are in the "strong-coupling" ($\frac{\mathcal{G}_h}{\kappa_0} \gtrsim 1$) regime regarding NM dissipations. Meanwhile, $\kappa_h$ can be kept below MHz considering that $Q_h \gtrsim 10^{10}$ has been realized [71–73]. The photon decay rate is analyzed in Section 3.3 of Ref. [23], where we show that $\kappa_h = 1$ MHz can be reached if two-photon absorption is avoided. We first fix $\kappa_0 = 0.1$ kHz and $\kappa_h = 1$ MHz. In Figure 3a, the time evolution of the NM population $n_0(t)$ is plotted for $\mathcal{G}_h$ in the weak-coupling ($\frac{\mathcal{G}_h}{\kappa_h} \ll 1$) regime. $n_0(t)$ monotonically decays with time, until reaching a steady-state value $n_0^{\text{steady}} = n_{\text{th}} \frac{\kappa_0 \kappa_h}{4\mathcal{G}_h^2 + \kappa_0 \kappa_h}$ (dashed line in Figure 3a, see Section 2.2 of Ref. [23] and Ref. [68]). With $\mathcal{G}_h = 10$ kHz (30 kHz), one has $\frac{n_0^{\text{steady}}}{n_{\text{th}}} \approx 0.20$ (0.027). Remarkably, the von Neumann entropy of the NM mode is suppressed as well. The entropy of the final steady state is close to that of a thermal state with a population of $n_0^{\text{steady}}$ (Section 2.3 of Ref. [23]). Then, we fix $\mathcal{G}_h = 10$ kHz. $n_0^{\text{steady}}$ as a function of $\kappa_0$ and $\kappa_h$ is shown in Figure 3b. A sizable cooling effect exists even when $\kappa_0 = 1$ kHz and $\kappa_h = 1$ MHz.

Next, we set $\mathcal{G}_h = 1$ MHz (Figure 3c) to demonstrate the laser cooling behavior in the strong-coupling regime. Note that this requires a strong pumping field $\mathcal{E}_p \sim 10^3$ MV/m, which can be challenging in practice. In this strong-coupling regime, there is a swap process between the NM and the $\omega_h$-photon with a frequency of $2\mathcal{G}_h$, while the total population ($n_0 + n_h$) drops with an envelope function $e^{-\bar{\kappa}t}$. The overall decay rate is $\bar{\kappa} \approx \frac{1}{2}(\kappa_0 + \kappa_h)$, because approximately the NM and the $\omega_h$-photon mode each exists for half of the time $t$ during the swap process. Finally, $n_0^{\text{steady}}$ reaches $\sim 10^{-4}$. Interestingly, when $\frac{\mathcal{G}_h}{\kappa_h} \gtrsim 1$, further increasing $\mathcal{G}_h$ does not improve the cooling effect. Instead, $n_0^{\text{steady}}$ is almost a constant $n_{\text{th}} \frac{\kappa_0}{\kappa_h}$ due to the back-heating effect (red curve in Figure 3d). Similar effects have also been observed in the case of optical cooling of phonons [74–76]. This limitation can be circumvented by $Q$-switching [30] (Section 2.4 of Ref. [23]). This would minimize the back-heating effect, and $n_0^{\text{steady}}$ can be further suppressed when $\frac{\mathcal{G}_h}{\kappa_h} \gtrsim 1$ (green curve in Figure 3d).

In summary, we introduce the ONQ effect, which can efficiently couple optical photons and nuclear spins. We demonstrate the laser cooling of NMs via the ONQ effect. The NM cooling process could be detected by monitoring the emission of cavity photons, and the occupation number reached could be measured by the dispersive frequency shift induced on a detuned anharmonic cavity (Section 3.4 of Ref. [23]). Since laser cooling suppresses both the population and the entropy of the NM mode, it could facilitate potential applications of nuclear spins, especially those based on the interface between nuclear spins and optical photons.

## Acknowledgment

This work was supported by an Office of Naval Research MURI through grant #N00014-17-1-2661 and Honda Research Institute USA, Inc., through grant #031807-00001. J.L. and A.B. also acknowledge support by DTRA (Award No. HDTRA1-20-2-0002) Interaction of Ionizing Radiation with Matter (IIRM) University Research Alliance (URA). The calculations in this work were performed in part on the Texas Advanced Computing Center (TACC) and MIT Engaging cluster. H.X. thanks Meihui Liu for help in figure production.